\documentclass[%
 superscriptaddress,
reprint,
 amsmath,amssymb,
 aps,longbibliography
]{revtex4-1}

\usepackage{graphicx}% Include figure files
\usepackage{dcolumn}% Align table columns on decimal point
\usepackage{bm}% bold math
\usepackage{hyperref}% add hypertext capabilities
\usepackage[mathlines]{lineno}% Enable numbering of text and display math
\usepackage[bottom]{footmisc}
\usepackage{booktabs}
\makeatletter
\renewcommand*{\@fnsymbol}[1]{\ensuremath{\ifcase#1\or *\or ** \or \ddagger\or
    \mathsection\or \mathparagraph\or \|\or **\or \dagger\dagger
    \or \ddagger\ddagger \else\@ctrerr\fi}}
\makeatother

\begin{document}
%\raggedbottomf
%\preprint{AIP/123-QED} 

\title{Hyperuniform scalar random fields for lensless, multispectral imaging systems}% Force line breaks with \\

\author{Yuyao Chen}
\affiliation{Department of Electrical and Computer Engineering and Photonics Center, Boston University, 8 Saint Mary’s Street, Boston, MA, 02215, USA}

\author{Wesley A. Britton}
\affiliation{Division of Materials Science and Engineering, Boston University, 15 St Mary’s St, Brookline, MA, 02246, USA}

\author{Luca Dal Negro}
\email[email:]{dalnegro@bu.edu}
\affiliation{Department of Electrical and Computer Engineering and Photonics Center, Boston University, 8 Saint Mary’s Street, Boston, MA, 02215, USA}
\affiliation{Division of Materials Science and Engineering, Boston University, 15 St Mary’s St, Brookline, MA, 02246, USA}
\affiliation{Department of Physics, Boston University, 590 Commonwealth Avenue, Boston, MA, 02215, USA}

%%%%%%%%%%%%%%%%%%%%%%%%%%%%%%%%%%%%%%%%%%%%%%%%%%%%%%%%%%%%%%%%%%%%%
%%%%%%%%%%%%%%%%%%%%%%%                  Abstract                      %%%%%%%%%%%%%%%%%%%%%%%%%%%%%
%%%%%%%%%%%%%%%%%%%%%%%%%%%%%%%%%%%%%%%%%%%%%%%%%%%%%%%%%%%%%%%%%%%%%
\begin{abstract}
We propose a novel framework for the systematic design of lensless imaging systems based on the hyperuniform random field solutions of nonlinear reaction-diffusion equations from pattern formation theory. Specifically, we introduce a new class of imaging point-spread-functions (PSFs) with enhanced isotropic behavior and controllable sparsity. We investigate the PSFs and the modulated transfer functions (MTFs) for a number of nonlinear models and demonstrate that two-phase isotropic random fields with hyperuniform disorder are ideally suited to construct imaging PSFs with improved performances compared to PSFs based on the Perlin noise. Additionally, we introduce a phase retrieval algorithm based on the non-paraxial Rayleigh-Sommerfeld diffraction theory and introduce diffractive phase plates with PSFs designed from hyperuniform random fields, called hyperuniform phase plates (HPPs). Finally, using high-fidelity object reconstruction, we demonstrate improved image quality using engineered HPPs across the visible range. The proposed framework is suitable for high-performance lensless imaging systems for on-chip microscopy and spectroscopy applications.
\end{abstract}

\maketitle
%%%%%%%%%%%%%%%%%%%%%%%%%%%%%%%%%%%%%%%%%%%%%%%%%%%%%%%%%%%%%%%%%%%%%
%%%%%%%%%%%%%%%%%%%%%%%                  Introduction                      %%%%%%%%%%%%%%%%%%%%%%%%%%%%%
%%%%%%%%%%%%%%%%%%%%%%%%%%%%%%%%%%%%%%%%%%%%%%%%%%%%%%%%%%%%%%%%%%%%%
Traditional imaging systems make use of lenses to focus the incoming radiation from a scene and form images onto a photodetector plane. However, the overall size and imaging quality produced by these systems is fundamentally limited by the numerical aperture (NA) of their lenses, even when considering most advanced metalens designs \cite{Khorasaninejadeaam8100,IPRNA}. In order to overcome this problem, lensless imaging systems were recently introduced that replace lenses with engineered diffractive optical elements (DOEs) placed in very close proximity to the detector plane, significantly reducing the system's volume \cite{DiffuseCam,Adams2017Single,boominathan2020phlatcam}. 
% For example, in typical lensless imaging implementations, a designed DOE of linear dimension $L\approx 7 ~mm$ is placed at a distance $d\approx 2~mm$ from the sensor plane and the image is reconstructed computationally based on the analysis of the diffracted intensity. 
Note that in this configuration the use of a single lens would significantly reduce the field of view due to severe aberrations at high numerical aperture \cite{Banerji:19,Liang2018Ultrahigh,IPRNA}. 
For these reasons, lensless cameras with largely reduced volumes already found numerous applications to on-chip microscopy, spectroscopy, 3D vision, and computer vision in autonomous vehicles \cite{DiffuseCam,Adams2017Single,boominathan2020phlatcam}. Besides, lensless cameras are cost-effective since they are fabricated in a scalable fashion using planar semiconductor technology \cite{boominathan2016lensless}. Unlike lens-based systems that produce tight focal spots characterized by strongly concentrated PSFs, lensless systems based on DOEs feature PSFs with larger supports \cite{DiffuseCam,Wu2020,boominathan2020phlatcam}. For example, a random diffuser or a Fresnel zone aperture (FZA) have been used  \cite{DiffuseCam,Wu2020}. Recently, a novel approach was introduced that leverages Gaussian random fields as suitable spatial structures for lensless PSFs engineering. In particular, lenseless PSFs have been designed based on the contour (i.e., the gradient map) of the Perlin noise and broader MTFs were demonstrated compared to random diffusers or FZAs \cite{boominathan2020phlatcam}. Robust phase retrieval algorithms are then used to compute the corresponding phase distributions of DOEs with the designed PSFs at specified distances from the detector plane. 
However, the Gaussian nature of the Perlin noise poses significant  challenges to further design progress. First, the Perlin noise is non-binary, requiring gradient-based edge detection in order to create binarized lensless PSFs. In the second place, the Perlin noise is non-isotropic, reducing the image quality. Finally, the  Perlin noise has only limited design parameters, preventing one to control the sparsity of lensless PSFs, which crucially determines the MTF bandwidths, once the sampling area has been fixed. 

In this paper, we address these important challenges by proposing a novel framework for the systematic design of lenseless PSFs based on  isotropic, non-Gaussian, and two-phase random fields. In particular, we design isotropic lensless PSFs directly from the solutions of stochastic reaction-diffusion partial differential equation (PDE) models under uniform random initial conditions. Specifically, we focus on the random field solutions of the Cahn-Hilliard (CH) \cite{cahn1958free}, Swift-Hohenberg (SH) \cite{swift1977hydrodynamic}, and Gray-Scott (GS)   \cite{gray1994chemical} nonlinear models. Importantly, these models naturally produce hyperuniform two-phase random fields that are stable upon thresholding \cite{ma2017random}, thus eliminating the need for gradient-based edge detection or alternative binarization techniques. We then introduce a phase retrieval algorithm based on the non-paraxial Rayleigh-Sommerfeld diffraction theory and obtain the 4-level discretized phase profiles of HPPs. Furthermore, we characterize their MTFs and demonstrate a $35\%$ enhancement of isotropicity with controlled sparsity compared to contour-PSFs based on the Perlin noise. 
Finally, we apply the high-fidelity alternating direction method of multipliers (ADMM) at different wavelengths and demonstrate multispectral object reconstruction using the designed HPPs. In particular, we obtain a $60\%$ improvement in the signal-to-background ratio (SBR) of the reconstructed objects using the CH model compared to Perlin PSFs.

\begin{figure}[h!]
\centering\includegraphics[width=\linewidth]{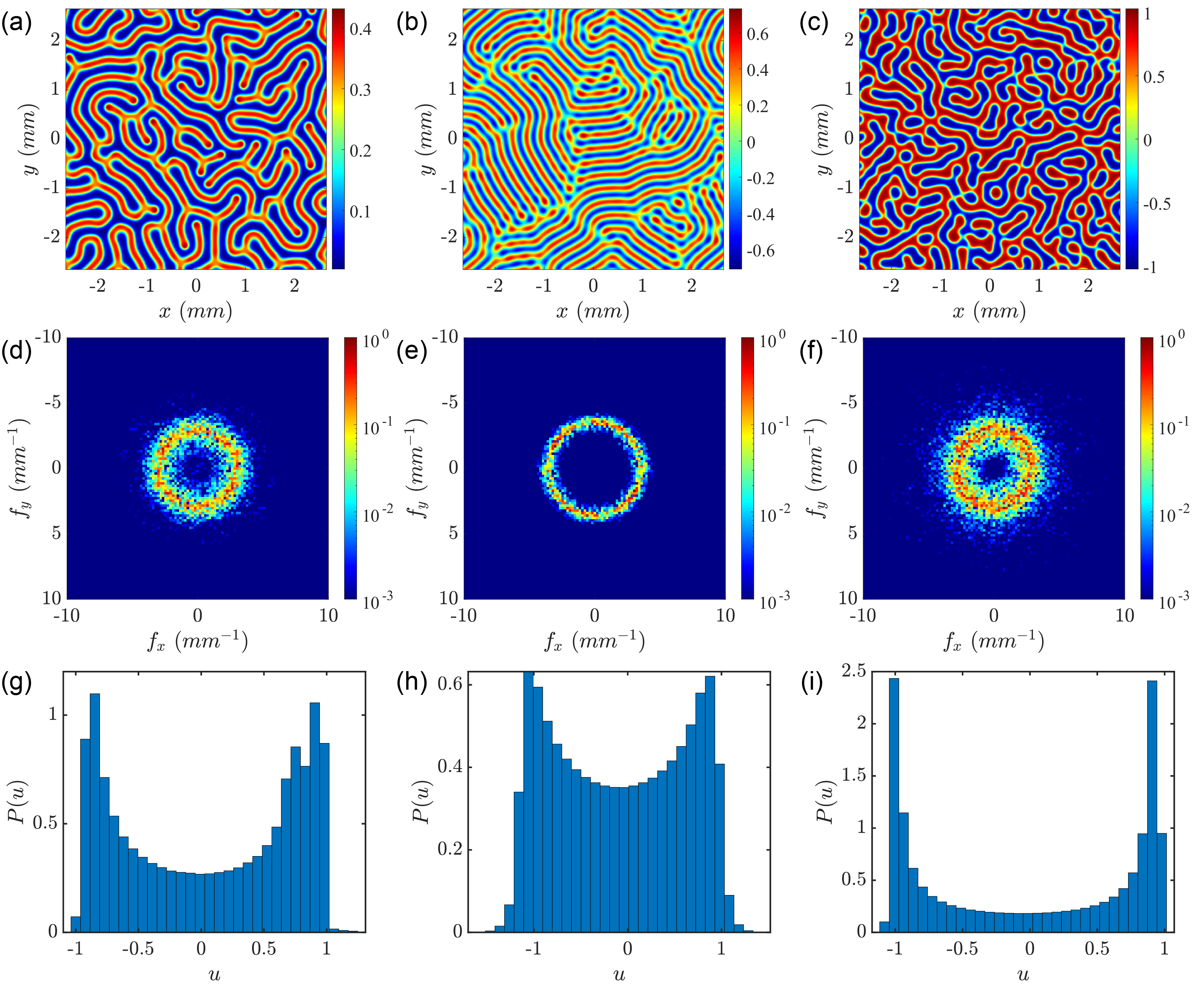}
\caption{(a-c) Realizations of the scalar random fields corresponding to the GS, SH, and CH models, respectively. (d-f) Spectral densities corresponding to the scalar random fields shown in panels (a-c), respectively. (g-i) Statistical distributions of the values of the random fields ensemble-averaged over 50 realizations of uniformly random initial conditions for the GS, SH, and CH models, respectively.}
\label{Fig1}
\end{figure}

All the lensless PSFs considered here can be regarded as scalar fields solutions of the general nonlinear reaction-diffusion equation:
\begin{equation}
\partial_{t} \boldsymbol{u}=\underline{\boldsymbol{D}} \nabla^{2} \boldsymbol{q}+\boldsymbol{R}(\boldsymbol{u})
\end{equation}
where $\underline{\boldsymbol{D}}$ is the diagonal diffusion coefficient matrix, and $\boldsymbol{R}$ is a nonlinear term that accounts for the local reactions \cite{lord2014introduction}. These nonlinear PDE models describe self-organized pattern/structure formation phenomena often encountered in chemistry, biology, and geophysics \cite{gray1994chemical}. Using a robust spectral method \cite{yoon2020fourier}, we solve the GS, SH, and CH models under uniformly random initial conditions and directly utilize their solutions for lensless PSF engineering. 
Typical realizations of the random field solutions for the three considered PDE models are shown in Fig. \ref{Fig1} (a-c), which clearly demonstrate their two-phase nature. To further characterize these solutions, we evaluate the corresponding spectral density functions, which are defined by the Fourier transform of the random field autocovariance function $\psi(\textbf{\textit{r}})=\langle[u(\textbf{\textit{x}}_{1})-\overline{u}(\textbf{\textit{x}}_{1})][u(\textbf{\textit{x}}_{2})-\overline{u}(\textbf{\textit{x}}_{2})]\rangle
$, where the angle bracket denotes the ensemble average \cite{lord2014introduction}. In Fig. \ref{Fig1} (d-f) we show the spectral density functions obtained according to the Wiener–Khinchin theorem \cite{lord2014introduction} from the power spectrum of the random field solutions in Fig. \ref{Fig1} (a-c). These functions are rotationally symmetric indicating that the random fields are isotropic, i.e., ${\psi}(\textit{\textbf{r}})={\psi}(r)$. Moreover, we notice that all the considered spectral density functions vanish at the origin of the frequency space, demonstrating their hyperuniform nature \cite{ma2017random,TORQUATO20181}.
%\begin{equation}
%    \lim _{|\mathbf{k}| \rightarrow 0} \tilde{\psi}(\mathbf{k})=0
%\end{equation}
%Here $\psi$ is the spectral density function. 
The two-phase character of the considered hyperuniform random fields is manifested in Fig. \ref{Fig1} (g-i) where we show the ensemble averaged results of the amplitude statistics over $50$ realizations for the GS, SH, and CH models, respectively. We scaled all the field values in the $[-1,+1]$ interval in order to directly compare the different models. 

\begin{figure}[t!]
\centering\includegraphics[width=\linewidth]{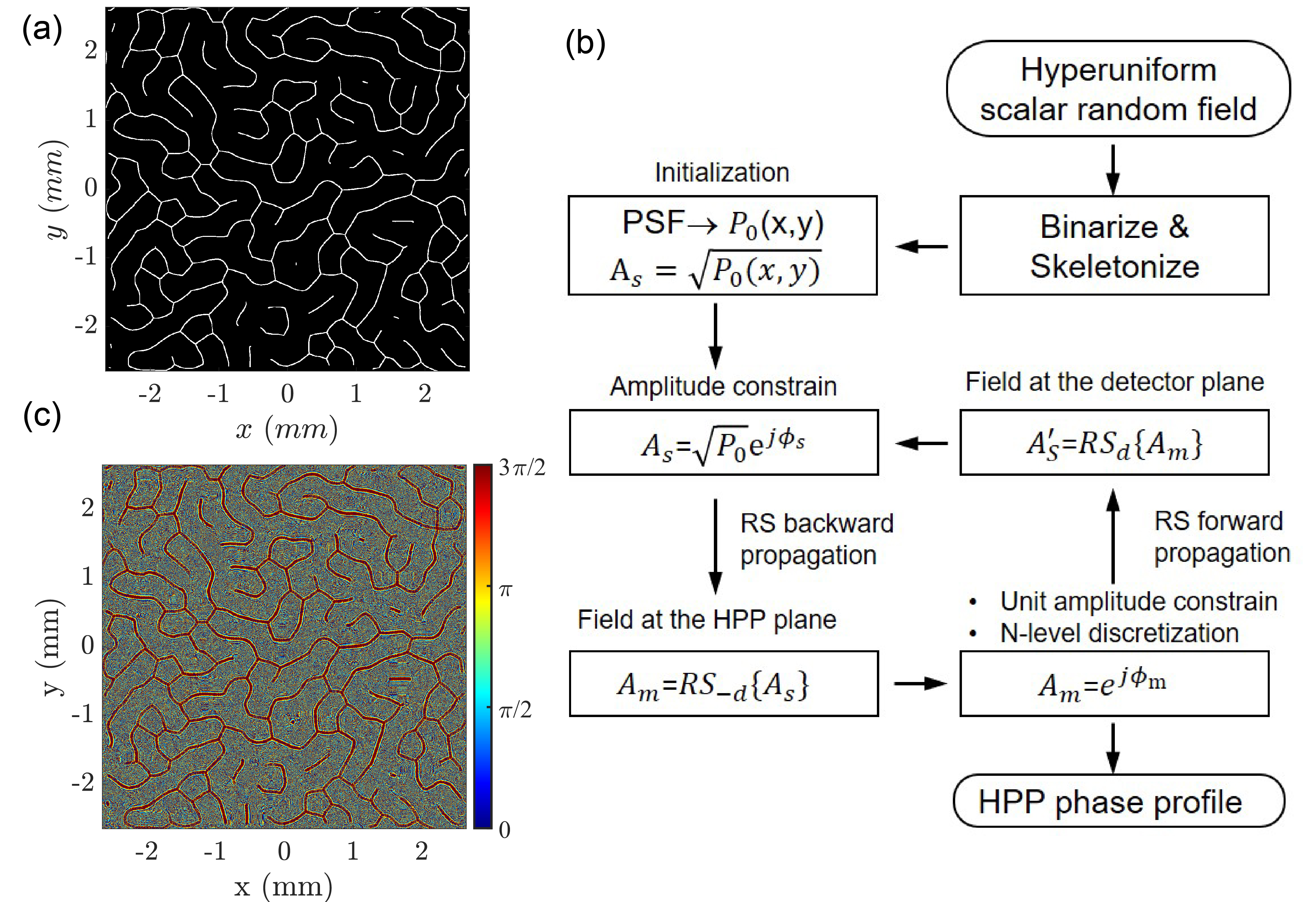}
\caption{(a) Representative PSF generated from the CH scalar random field shown in Fig. \ref{Fig1} (c). (b) Phase retrieval algorithm used to create the 4-level discretized phase mask shown in (c).}
\label{Fig2}
\end{figure}

In order to create imaging PSFs starting from the hyperuniform random fields we computed their topological skeletons, resulting in binary contour PSFs lines with the same topology of the original field (i.e., with the same Euler number) \cite{LEE1994462}. This procedure is generally needed to obtain sparse binary PSFs with increased MTF bandwidths that compensate for the limited bit depth of the detector \cite{boominathan2020phlatcam}. Since in our case the scalar random fields are already two-phase systems, the binarized PSF skeletons maintain the underlying isotropic behavior. As an example we display in Fig. \ref{Fig2} (a) the lensless PSF obtained from the skeleton of the random field realization displayed in Fig. \ref{Fig1} (c). 
We now utilize a Gerchberg-
Saxton (GS) phase retrieval algorithm \cite{gerchberg1972practical} coupled to the Rayleigh-Sommerfeld first integral in order to propagate forward to the detector plane, where the lensless PSF is obtained, and backward to the mask plane, where the HPP is located. The Rayleigh-Sommerfeld (RS) first integral is defined by \cite{goodman2005introduction,britton2020compact,britton2020phase,Chen:20,Chen:21}:
\begin{eqnarray}\label{RS equation}
    A_{s}\left(x^\prime,y^\prime\right)= A_{m}\left(x,y\right) * h(x,y;d,k)\\
    h(x,y;d,k)=\frac{1}{2\pi}\frac{d}{r}  \left(\frac{1}{r}-jk\right)\frac{e^{jkr}}{r}.
\end{eqnarray}
where $*$ denotes two-dimensional space convolution, $A_{m}$, $A_{s}$ are the transverse field distributions in the HPP and detector planes with coordinates $(x,y)$ and $(x^\prime,y^\prime)$, respectively. Moreover, $k$ is the incident wave number and $r=\sqrt{x^2+y^2+d^2}$, where $d$ is the distance between HPP and detector plane. We denote the forward propagation by a distance $d$ as $A_{s}=RS_{d}\{A_{m}\}$ in Fig. \ref{Fig2} (b), where the algorithm is presented. The backward propagation from the detector plane to the mask plane is obtained by inverting the sign of the distance $d$. We implement iteratively the RS forward and backward propagation between the two planes and apply the constraints that the $A_m$ intensity is unity and its phase is 4-level discretized while the $A_s$ coincides with the lensless PSF. Using this procedure we obtain the HPP phase profiles on the sensor plane that produce the desired PSF at distance $d$ for the wavelength $\lambda$. 
To illustrate the application of the presented methodology we select the PSF displayed in Fig. \ref{Fig1} (c) and we perform the phase retrieval by considering $d=2~mm$ and $\lambda=532.8~nm$ within a square aperture of length $L=5.3~mm$. We show the resulting HPP phase profile in Fig. \ref{Fig2} (c). The spatial resolution is $\Delta x =4~{\mu}m$, which is suitable for fabrication using scalable photolithography of 4-level diffractive elements \cite{britton2020compact,britton2020phase}. However, the retrieved HPP phase profile can also be readily implemented using planar metasurface technology \cite{Khorasaninejadeaam8100}, enabling advanced applications of HPP-based lensless imaging systems.

\begin{figure}[t!]
\centering\includegraphics[width=\linewidth]{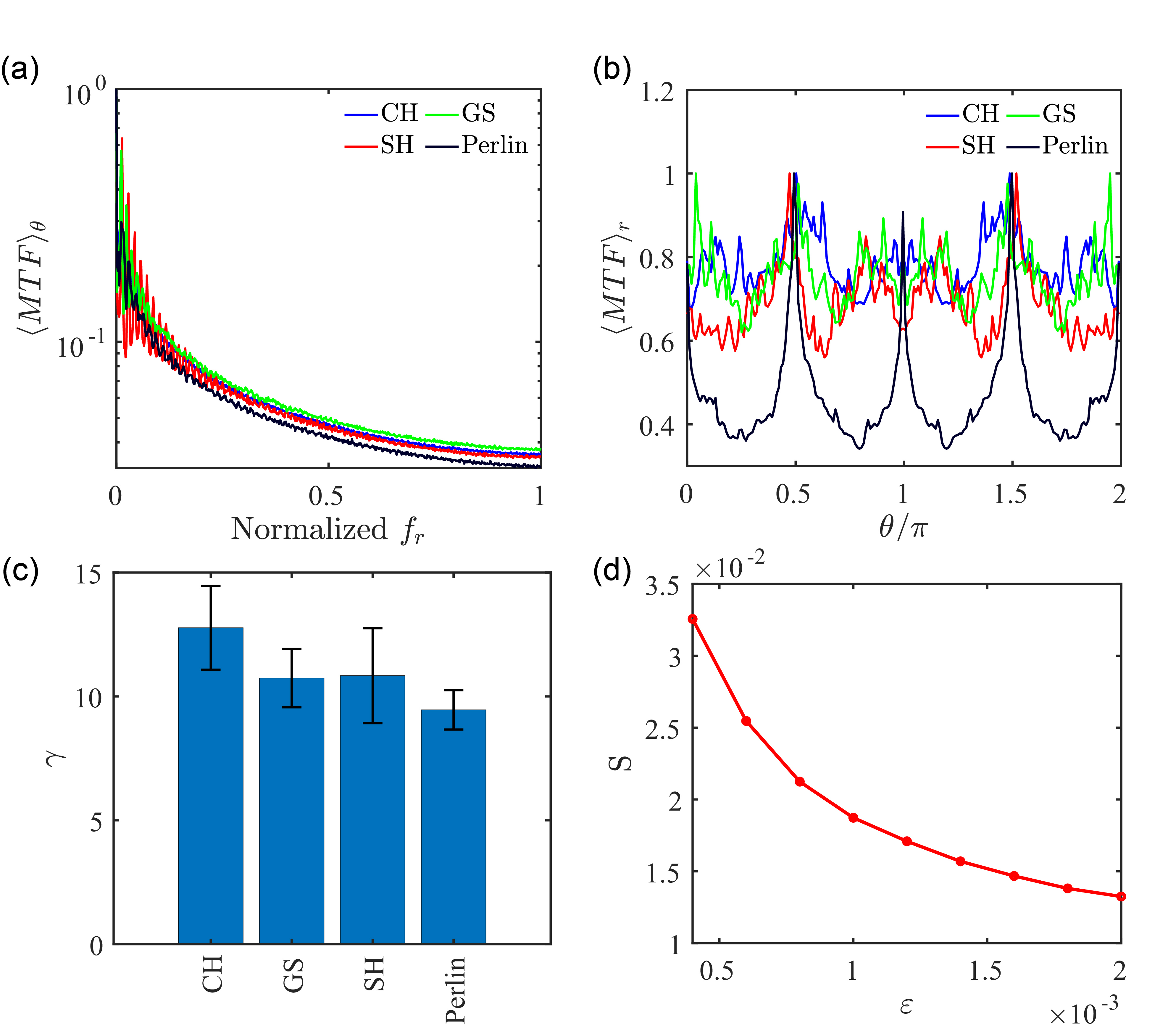}
\caption{(a) Azimuthally averaged and (b) radially averaged MTFs for the different lensless PSFs indicated in the legend. (c) Isotropic parameter $\gamma$ computed for different lensless PSFs considering $50$ realizations of uniformly random initial conditions. (d) Sparsity of the CH-based PSF with respect to the model parameter $\varepsilon$ in the CH equation.}
\label{Fig3}
\end{figure}

We now characterize the MTFs of the proposed lensless systems, which are obtained from the Fourier spectral amplitude of the imaging PSFs. In particular, in Fig. \ref{Fig3} (a) we show the azimuthally-averaged MTFs corresponding to PSFs with the same sparsity. The sparsity $S$ of a binary PSF is defined by the fraction of non-zero pixels within its square support area.
%\begin{equation}
%    S=\frac{ \int\int PSF da}{A_{max}}
%\end{equation}
%where the $A_{max}=L^2$ is the largest support area of the PSF. 
It has been shown that the PSF sparsity directly determines the MTF bandwidth \cite{boominathan2020phlatcam}. This is consistent with our results shown in Fig. \ref{Fig3} (a) where different PSFs with identical sparsity approximately display the same MTF bandwidth.
%since a binary PSF with lots of zero-values (small $S$) can reduce its DC component in the MTF and thus increase the MTF bandwidth 
Therefore, it is crucial for lensless PSF engineering to be able to flexibly control the PSF sparsity. Moreover, in Fig. \ref{Fig3} (b) we plot the radially-averaged MTF with respect to the azimuthal angle in the frequency plane for the PSF models specified in the legend. Given that all the investigated PSFs have approximately the same sparsity, it follows that the ones derived from the considered hyperuniform random fields feature smaller MTFs azimuthal variations compared to the ones based on Perlin noise. We further quantify the isotropic behavior of the MTFs associated to the considered lensless PSFs by defining their \textit{isotropic parameter} $\gamma$ as: 
\begin{equation}
\gamma=1/\sigma_\theta\left(\langle MTF \rangle_r\right)
\end{equation}
where the $\sigma_\theta$ denotes the standard deviation with respect to the angle $\theta$ on the frequency plane. We evaluate $\gamma$ over $50$ realizations of the random fields and present our results in Fig. \ref{Fig3} (c). The error bars are derived from the standard deviation over different realizations. The results demonstrate that the lensless PSFs designed from the proposed random fields are more isotropic compared to the Perlin noise PSF. In particular, the more pronounced bimodal amplitude statistics of CH random field solutions shown in Fig. \ref{Fig1} (i) translates into a $35\%$ increase in the isotropic parameter $\gamma$ of the PSF compared to the Perlin noise counterpart. More isotropic PSFs/MTFs can better capture spatial frequencies at different angles with a similar responses, which leads to higher-quality objects reconstruction, as we will discuss later. 

An additional advantage of the proposed approach is that we can achieve systematically control the  sparsity of the PSFs by tuning the parameters of the corresponding PDE models. For example, the CH equation for the random field $u$ is explicitly given by \cite{cahn1958free}:
\begin{equation}
\frac{\partial u(\mathbf{x}, t)}{\partial t}=\Delta\left(u^3 -u -\varepsilon^{2} \Delta u\right)
\end{equation}
where the parameter $\varepsilon$  sets the width of the transition regions between different phases in the obtained random field solutions. The effect of tuning the parameter $\varepsilon$ of CH equation on the sparsity of the corresponding PSF is shown in Fig.\ref{Fig3} (d), where we observe that the sparsity $S$ decreases monotonically when increasing $\varepsilon$. We note that sparsity control can also be achieved for PSFs derived from the GS and SH models, which in fact are specified by two parameters \cite{swift1977hydrodynamic,gray1994chemical}. Therefore, the engineering of lensless PSFs from the nonlinear stochastic PDEs of pattern formation theory provides several degrees of freedom to control both the PSF and MTF for a fixed sampling area, which cannot be achieved with simpler Gaussian noise models. 

To further characterize the imaging capability of the CH-based PSF, we perform high-fidelity object reconstruction simulations that recover an object from the diffracted intensity captured by the detector. We consider to the incoherent diffraction-limited image formation model \cite{boominathan2020phlatcam,DiffuseCam}:
\begin{equation}
     I(x,y)= PSF(x,y) * O(x,y)
\end{equation}
where $I$ is the captured intensity by the detector, $*$ denotes the two-dimensional spatial convolution and $O$ is the imaged object. The object reconstruction problem can then be formulated as an optimization problem within the ADMM algorithm  \cite{boyd2011distributed}. This is a commonly used iterative method for high-fidelity object reconstruction in lensless computational imaging \cite{DiffuseCam,boominathan2020phlatcam}. Specifically, we solve the following optimization problem:
\begin{equation}
\hat{O}=\arg \min _{O \geq 0} \frac{1}{2}\|I- \mathbf{P} O\|^{2}_{2}+\sigma\|\Psi(O)\|_{1}
\end{equation}
where $\|\cdot\|_1$ and $\|\cdot\|_2$ are the $\ell_1$ and $\ell_2$ norms, respectively, $\sigma$ is the weighting factor for the regularization term. The symbol $\Psi$ denotes the 2D gradient operator and $\mathbf{P}$ represents the PSF convolution. 
We then use variable splitting as follows:
\begin{equation}
\hat{O} =\arg \min _{w \geq 0, z, O_j} \frac{1}{2}\|\mathbf{I}- O\|^{2}_{2}+\sigma\|z\|_{1}
\end{equation}
where we used the notations that $z=\Psi x$ and $w=O$. 
Further details on the implementation steps of this method can be found in the reference \cite{boyd2011distributed}. 
%Then the update for the iteration at step $k$ are 
%\begin{align}
%z^{k+1} & \leftarrow \mathcal{S}_{\frac{\sigma}{\mu_{z}}}\left(\Psi O+\rho_{z}^{k} / \mu_{z}\right) \\
%w^{k+1} & \leftarrow \max \left(O+\rho_{w}^{k} / \mu_{w}, 0\right) \\
%O_j^{k+1} & \leftarrow\left(\mathbf{P}^{T} \mathbf{P}+\mu_{z} \Psi^{T} \Psi+\mu_{w} I\right)^{-1} r^{k} \\ 
%\rho_{z}^{k+1}  & \leftarrow \rho_{z}^{k} +  \mu_z (\Psi O^{k+1} - z^{k+1})\\ 
%\rho_{w}^{k+1}  & \leftarrow \rho_{w}^{k} +  \mu_w (\Psi O^{k+1} - w^{k+1}) \end{align}
%where
%\begin{equation}
%r^{k}=H^{T} \mathbf{b}+\Psi^{T}\left(\mu_{z} z^{k+1}-\rho_{z}^{k}\right)+\mu_{w} w^{k+1}-\rho_{w}^{k}
%\end{equation}
%$\mathcal{S}_{\frac{\gamma}{\mu_{z}}}$ is the soft-thresholding operator with threshold value $\mu$. The symbols $\rho_z$ and $\rho_w$ denote the Lagrangian multipliers for $z=\Psi x$ and $w=O$, respectively. The $\mu_w$ and $\mu_z$ are the penalty parameters for $\rho_w$ and $\rho_z$, respectively. 
\begin{figure}[t!]
\centering\includegraphics[width=\linewidth]{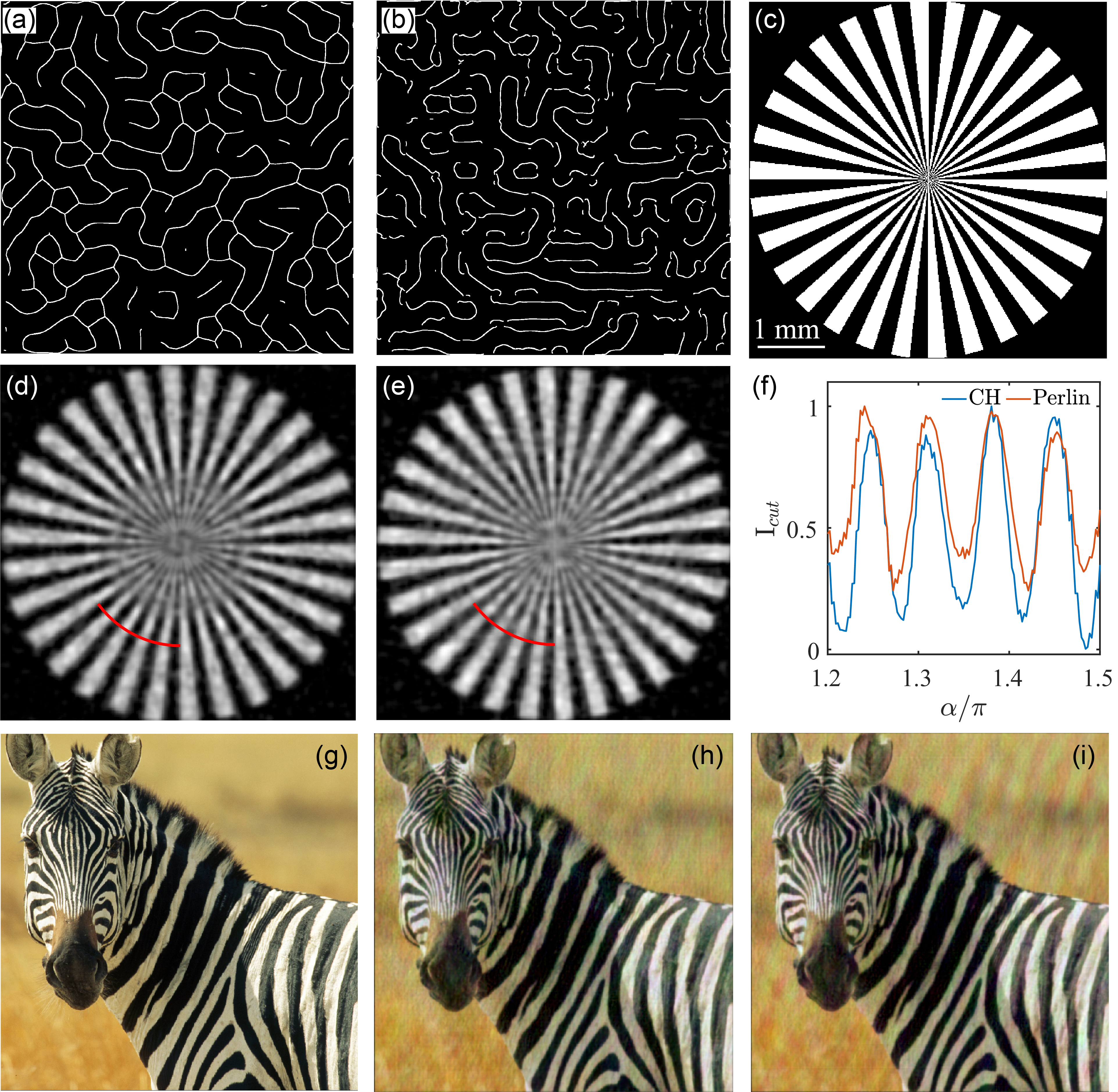}
\caption{ADMM object reconstruction using the CH-based PSF and Perlin-based PSF shown in (a) and (b), respectively. (c)  Original object (star target positioned $2~mm$ in front of the HPP) used for the reconstruction. (d,e) Reconstructed objects using CH- and Perlin-based PSFs, respectively. (f) Intensity profiles along the red lines indicated in panels (d) and (e). (g) Multi-spectral scene. The analyzed scene has a side length equal to $5.3~mm$ and the is placed $2~mm$ away from the lensless phase mask. Reconstructed multi-spectral scenes using the CH-based (h) and the Perlin-based PSF (i).}

\label{Fig4}
\end{figure}

We show our reconstructed results  using the star target object in Fig \ref{Fig4} (c). This object has a high degree of rotational symmetry and is suitable to better illustrate the behavior of the PSF at different angles. We display in Fig. \ref{Fig4} (d) and (e) the reconstructed object using the ADMM algorithm for the CH- and Perlin-based PSFs shown in Fig. \ref{Fig4} (a) and (b), respectively. In Figure \ref{Fig4} (f) we compare the intensity profiles of the reconstructed objects along the red trajectories visible in panels (d) and (e). The results clearly show that the CH-based PSF reconstructs the object with a larger contrast. To fully characterize the image quality, we compute the signal-to-background ratio (SBR) evaluated as the azimuthal and radial average of the intensity on the reconstructed object divided the one of the background region outside the object \cite{Sandison:94}. The SBRs values obtained for the reconstructed objects using the considered PSFs are $9.5$ (CH) and $5.9$ (Perlin). Therefore, a $60\%$ enhancement in the SBR of the CH-based HPP with respect to the Perlin PSF is achieved. On the other hand, the SBR enhancement values obtained for the GS and SH models are $17\%$ and $14\%$. We attribute this difference to the more isotropic nature of the CH-based PSF. Finally, we perform multispectral imaging simulations of the synthetic scene shown in Fig. \ref{Fig4} (g). We consider a 4-level discretized HPP ($d=2~mm$ and $\lambda_0=533~nm$) and compute the PSFs at additional wavelengths using Eq. \ref{RS equation}. The imaging simulations are then performed by reconstructing the synthetic scene at each wavelength using the corresponding PSFs. Specifically, we show the multispectral reconstruction using the wavelengths $\lambda_0$, $\lambda_1=420~nm$, and $\lambda_2=580~nm$ and we show the results in Fig. \ref{Fig4} (h) and (i) for the CH- and Perlin-based PSFs, respectively. The quality of these reconstructions can be measured by the SSIM parameter \cite{wang2004image}, which is equal to $0.83$ (CH) and $0.8$ (Perlin) for the results shown in Fig. \ref{Fig4} (h) and (i), respectively. The reconstruction based on the designed HPP demonstrates better image quality. However, this may vary for different objects and the SSIM parameter should only be considered a qualitative metric \cite{boominathan2020phlatcam}.

In conclusion, we proposed a novel framework for the systematic design of lensless imaging PSFs with controlled sparsity based on isotropic random fields from the hyperuniform solutions of nonlinear reaction-diffusion equations. 
We presented a phase retrieval algorithm based on Rayleigh-Sommerfeld diffraction to introduce novel diffractive phase plates that can be experimentally demonstrated using multi-level diffractive elements or planar metasurfaces, enabling advanced applications to lensless imaging systems. We characterized their MTFs and demonstrated a $35\%$ enhancement in the isotropic parameter compared to Perlin noise PSFs. We further applied high-fidelity ADMM object reconstruction and achieved $60\%$ SBR enhancement compared to Perlin PSFs. Finally, we demonstrated multispectral imaging with improved quality, which could enable high-performance applications of lensless microscopy and spectroscopy. 

\clearpage

\section*{Funding Information}
Army Research Laboratory (W911NF-12-2-0023); National Science Foundation (ECCS-2015700).

\section*{Disclosures}
The authors declare no conflicts of interest.

\bibliography{sample}

\end{document}